\documentclass[aps,prb,nofootinbib,twocolumn,showpacs,letterpaper,tighten,float,superscriptaddress,reprint]{revtex4-1}

\usepackage{amssymb}
\usepackage{amsbsy}
\usepackage{amsmath}
\usepackage{amsfonts}
\usepackage{mathrsfs}
\usepackage{epsfig}
\usepackage{graphicx}
\usepackage{array}
\usepackage{textcomp}
\usepackage{color}
\usepackage{braket}
\usepackage{bm,hyperref}
\usepackage[justification=raggedright,font=small]{caption}
\usepackage[titletoc,toc,title]{appendix}
\usepackage{ulem}
\usepackage[mathscr]{euscript}
\usepackage{mathtools}
\usepackage{dsfont}

\usepackage[labelformat=simple]{subcaption}

\definecolor{myblue}{rgb}{.93, .93, 1}

\definecolor{darkgreen}{rgb}{0,0.7,0}

\newcommand{\beq}{\begin{equation}}
\newcommand{\eeq}{\end{equation}}

\newcommand{\bpm}{\begin{pmatrix}}
\newcommand{\epm}{\end{pmatrix}}
\newcommand{\bmm}{\begin{matrix}}
\newcommand{\emm}{\end{matrix}}

\newcommand{\br}{\bf{r}}
\newcommand{\hphi}{\hat{\phi}}

\newcommand{\combinationNumber}[2]{\left(\begin{matrix}
#1  \\
#2  \\
\end{matrix}\right)}

\graphicspath{{./Figures/}}

\begin{document}

\author{Huan He}
\affiliation{
Department of Physics,
Princeton University, NJ 08544, USA
}
\author{Yizhi You}
\affiliation{
Princeton Center for Theoretical Science,
Princeton University, NJ 08544, USA
}
\author{Abhinav Prem}
\thanks{Corresponding author: aprem@princeton.edu}
\affiliation{
Princeton Center for Theoretical Science,
Princeton University, NJ 08544, USA
}

\title{Lieb-Schultz-Mattis type constraints on Fractonic Matter}
\date{\today}

\begin{abstract}

The Lieb-Schultz-Mattis (LSM) theorem and its descendants impose strong constraints on the low-energy behavior of interacting quantum systems. In this paper, we formulate LSM-type constraints for lattice translation invariant systems with generalized U(1) symmetries which have recently appeared in the context of fracton phases: U(1) polynomial shift and subsystem symmetries. Starting with a generic interacting system with conserved dipole moment, we examine the conditions under which it supports a symmetric, gapped, and non-degenerate ground state, which we find requires that both the filling fraction and the bulk charge polarization take integer-values. Similar constraints are derived for systems with higher moment conservation laws or subsystem symmetries, in addition to lower bounds on the ground state degeneracy when certain conditions are violated. Finally, we discuss the mapping between LSM-type constraints for subsystem symmetries and the anomalous symmetry action at boundaries of subsystem symmetric topological (SSPT) states.

\end{abstract}

\maketitle


\section{Introduction}
\label{sec:intro}

Determining the nature of the ground state of a quantum many-body system, given its underlying microscopic degrees of freedom and the Hamiltonian governing their dynamics, is a problem of fundamental import but is typically rendered intractable by analytical or numerical approaches in the presence of interactions. Nevertheless, in some cases it is possible to prove that the low-energy behavior of a (possibly interacting) system is strongly constrained by certain properties of the microscopic input---specifically, by the manner in which symmetries act on the many-body Hilbert space. In lieu of exact analytical solutions, it is hence highly desirable to establish rigorous results which place non-trivial constraints on the many-body ground state(s) of generic interacting systems.

The celebrated Lieb-Schultz-Mattis (LSM) theorem~\cite{LSM1961}, as well as its higher dimensional generalizations by Oshikawa~\cite{oshikawa1} and Hastings~\cite{hastings2004}, provides an archetypal example of ultraviolet data non-trivially constraining the universal infrared behavior of lattice spin-systems. In the thermodynamic limit, the theorem forbids a symmetric, gapped, and unique ground state in any translation invariant spin system with an odd number of spin-$1/2$'s per unit cell. Since a short-range-entangled (SRE) ground state preserving both translation and SO(3) symmetries is ruled out, either some symmetry is spontaneously broken, the ground state is gapless, or (in $d \geq 2$ spatial dimensions) the ground state has a non-trivial degeneracy on the $d$-torus. The latter corresponds to a topologically ordered phase (such as a topological quantum spin liquid), which is gapped, symmetric, and hosts fractionalized excitations with non-trivial mutual statistics. LSM-type theorems thus harbor significant implications for strongly correlated systems, including quantum spin liquids~\cite{lee2008,savaryreview} and deconfined quantum critical points~\cite{senthilSPTreview,song2018spinon,lee2019signatures,ning2019fractionalization,zou2018bulk}.

Given the power of the LSM theorem, there has been much progress in extending its scope to various other settings and in establishing close relationships of  these LSM-type constraints with the theory of symmetry protected topological (SPT) phases, as well as their implications for symmetry enriched topological (SET) phases~\cite{oshikawa2,oshikawasenthil,hastings2005,roy2012,sid2013,watanabe2015,watanabe2016,furuya2017,po2017,cho2017,huang2017,jian2018,metlitski2018,thorngren2018,cheng2019fermionic,jiang2019generalized,else2019topological,zaletel2015,cheng2016,hsieh2016,yang2018,bultinck2018,lu2017,lu2017b,huang2017,qi2017,qi2018,unsal2018,watanabe2018,tanizaki2018,shiozaki2018,song2018,yao2019}. For example, the LSM theorem has been generalized to other symmetry groups, including internal symmetries besides SO(3) spin-rotation and space group symmetries beyond lattice translation~\cite{roy2012,sid2013,watanabe2015,watanabe2016,furuya2017,po2017,cho2017,huang2017,jian2018,metlitski2018,thorngren2018,cheng2019fermionic,jiang2019generalized,else2019topological}, as well as to higher-form U(1) symmetries~\cite{ryuLSM}.

Recently discovered fracton phases of matter~\cite{chamon,haah,yoshida,fracton1,fracton2}, which are long-range-entangled states transcending the conventional TQFT paradigm, have garnered significant recent interest~\cite{williamson,slagle1,prem,han,hsieh,regnault2,leomichael,shirleygeneral,albert,shirleyfrac,cagenet,pai2,yizhilego,slaglesmn,premgauging,bulmashgauging} by virtue of their striking properties (see Ref.~[\onlinecite{fractonreview}] for a review). Chief amongst these is the restricted mobility of topological excitations, with fundamental charges fully immobile while the dynamics of composite multipole objects are constrained to lie along lower-dimensional manifolds. Such strict constraints can arise from gauging certain global symmetries: polynomial shift symmetries and subsystem symmetries. The former are associated with higher moment conservation laws (such as dipole moment) in systems with a conserved U(1) charge and, when gauged, lead to fractonic gauge theories~\cite{sub,genem,prem2,han3,pai,bulmash,symmetric,gromov,bulmash2,fractongauge,albertgauge,williamsonSET,gromov2019,you2019fractonic,embeddon,wangnonabelian}. Subsystem symmetries correspond to conservation laws whose associated charges are conserved along sub-manifolds (lines, planes, or fractals) and can lead to new SPT states, dubbed subsystem SPTs (SSPTs), many of which are dual to fracton phases through a generalized gauging procedure~\cite{fracton2,williamson,xumoore,yizhi1,yizhi2,devakulfractal,strongsspt,shirleygauging,shirleytwisted,devakulstrong}. By now, both classes of symmetries have begun attracting attention independent of their connection to fractons~\cite{spurious,devakulMBQC,stephen2018,albertespec,you2019multipole,you2019higher,hughes2019dipole}.

In this work, we put forth LSM-type constraints for lattice-translation invariant quantum many-body systems with generalized U(1) symmetries, namely polynomial shift and subsystem symmetries. In particular, we show that the interplay of dipole moment conservation and lattice translation symmetry forbids a symmetric, gapped, and non-degenerate state when either the charge density is fractional or when the charge density is integer-valued but the bulk displays a fractional charge polarization (equivalently, a fractional dipole moment per unit length). We also briefly comment on the implications of the latter constraint for topological dipole insulators~\cite{hughes2019dipole} as well as its extension to systems with higher moment conservation laws, details of which are presented in the Appendix. We then generalize these arguments to systems with U(1) subsystem symmetries and relate the resulting constraints to boundaries of SSPT states. We expect that our results will aid in the search and discovery of novel spin liquids with emergent fractonic behavior.


\section{Constraints for dipole moment conserving systems}
\label{sec:multipole}

As initially pointed out in Ref.~[\onlinecite{sub}], much of the fractonic phenomenology can be captured by symmetric tensor gauge theories with conserved higher moments, such as dipole moment. Later, it was realized by Gromov~\cite{gromov2019} that the symmetries responsible for such higher moment conservation laws are \textit{polynomial shift symmetries} which, upon gauging, lead to fractonic gauge theories \textit{i.e.,} gauge theories whose fundamental charges and composites thereof have restricted mobility. In this section, we briefly review the usual LSM-Oshikawa-Hastings (LSMOH) theorem for lattice translation invariant systems with a conserved U(1) global symmetry, following which we discuss $d=2$ systems which also conserve the global dipole moment. This conservation law stems from the simplest non-trivial polynomial shift symmetry: a linear shift symmetry. Generalizations to arbitrary polynomial shift symmetries are also discussed briefly, with technical details relegated to the Appendix.

\subsection{Brief review of the LSMOH theorem}
\label{sec:review}

The LSMOH theorem~\cite{LSM1961,oshikawa1,hastings2004} imposes non-trivial constraints on the ground state(s) of quantum many-body systems based purely on their global symmetries and commensurability (or lack thereof) with the underlying lattice, an example of the ultraviolet dictating the infrared behavior. We illustrate this through the simple example of a $d=2$ quantum many-body system comprised of a single particle species and invariant under both lattice translation and a global U(1) symmetry, the latter of which corresponds to particle number conservation. Since the particles need not carry real electric charge, we assign a fictitious charge $e$ to each particle, which couples to a fictitious externally specified field. 

The system is defined on an $L_x \times L_y$ lattice, with the length $L_x$ measured in terms of $a$, the lattice spacing in the $x$-direction. The boundary conditions are taken to be periodic (open) in the $x$ ($y$) direction, such that the system is defined on a cylinder. Hence, the Hamiltonian $[\hat{H},\hat{T}_x] = 0$, where the operator $\hat{T}_x$ translates the system by $a$ along the $x$-direction. On the lattice, elements of the global U(1) symmetry group are given by
\beq
\hat{D}(\theta)=\exp \left( i \theta \sum_{\br} \hat{n}_{\br} \right) \, ,
\eeq
where $\theta$ denotes the global phase shift of the U(1) symmetry and $\hat{n}_{\br}$ is the number operator on the $\mathbf{r}^{th}$ site, with the summation taken over all sites. Now, the operator $\hat{Q} = \sum_{\br} \hat{n}_{\br}$ measures the total charge of the system, with the charge density (equivalently, filling fraction) thus defined as $\nu = Q/(L_x L_y)$. Note that $\nu = p/q$ is always a rational number, with $p$ and $q$ coprime. 

Since the Hamiltonian commutes with $\hat{T}_x$, we can always choose the ground state $\ket{\Psi}$ such that it is also an eigenstate of $\hat{T}_x$ with eigenvalue $t_x$. Following Laughlin's flux threading argument~\cite{laughlin}, we imagine starting in a many-body ground state and adiabatically inserting a unit flux quantum $\Phi_0 = 2\pi$ through the hole of the cylinder (in units where $e = \hbar = c = 1$). Since there is a trivial Aharonov-Bohm effect during this procedure, the energy spectrum remains unchanged; indeed, the Hamiltonian with unit flux quantum $\hat{H}(\Phi_0)$ maps onto the original one $\hat{H}(0)$ through the large gauge transformation
\beq
\label{eq.U1LargeGaugeTransformation}
\hat{D}_x = \exp \left( \frac{2\pi i}{L_x} \sum_{\br} x_{\br} \hat{n}_{\br} \right) \, ,
\eeq
where $x_{\br}$ is the coordinate of the $\mathbf{r}^{th}$ site in the $x$-direction. This large gauge transformation maps the Hamiltonian after the flux insertion back to the original Hamiltonian as $\hat{D}_x\hat{H}(\Phi_0) \hat{D}_x^\dagger=\hat{H}(0)$. 

Unlike the Hamiltonian, the ground state $\ket{\Psi}$ evolves into some distinct state $\ket{\Phi}$ after flux insertion, which nevertheless has the same eigenvalue $t_x$ as the original state $\ket{\Psi}$ since $\hat{H}$ commutes with $\hat{T}_x$ throughout the adiabatic process. After applying the large gauge transformation, Eq.~\eqref{eq.U1LargeGaugeTransformation}, the state $\ket{\Phi}$ is then mapped onto $\hat{D}_x \ket{\Phi}$. Thus, after the adiabatic flux insertion and large gauge transformation, the Hamiltonian returns to itself but the ground state $\ket{\Psi}$ is mapped onto the state $\hat{D}_x \ket{\Phi}$. Since the adiabatic theorem ensures that the ground state remains in the ground state manifold after flux insertion~\cite{laughlin,tao} and since $\hat{D}_x$ is a large gauge transformation, the state $\hat{D}_x \ket{\Phi}$ is also a ground state of the original Hamiltonian, albeit with a possibly different $\hat{T}_x$ eigenvalue than $t_x$. However, we note the non-trivial commutation relation between $\hat{T}_x$ and $\hat{D}_x$:
\beq
\begin{split}
\hat{T}_x \hat{D}_x \hat{T}^{-1}_x 
=& \hat{D}_x \exp \left(-\frac{2\pi i}{L_x} \sum_{\br} \hat{n}_{\br} \right) \, .
\end{split}
\eeq
In terms of $\nu$, the above commutation relation becomes
\beq\label{eq.commutationTDin2D}
\begin{split}
\hat{T}_x \hat{D}_x=\exp \left( -2\pi i \nu L_y \right) \hat{D}_x \hat{T}_x \, .
\end{split}
\eeq
We thus see that the state $\hat{D}_x \ket{\Phi}$ has a $\hat{T}_x$ eigenvalue $t_x - 2 \pi \nu L_y$, which implies that $\hat{D}_x \ket{\Phi}$ is orthogonal to both $\ket{\Psi}$ and $\ket{\Phi}$ when $L_y$ and $q$ are coprime. Thus, in the presence of a finite excitation gap, a fractional charge density (or fractional filling factor) $\nu = p/q$ implies that the ground state manifold is at least $q$-fold degenerate, which forbids a symmetric, gapped, and non-degenerate ground state. Conversely, if the gapped system has a unique ground state and the ground state preserves both U(1) and translation symmetries, then the charge density $\nu$ is necessarily an integer. Note that picking a special choice of $L_y$ is reasonable since we are interested in the thermodynamic limit in the preceding argument.

The above argument easily generalizes to arbitrary dimensions, with only the lattice details requiring modification. Specifically, in $d$-dimensions, we consider a lattice with periodic boundary conditions (PBC) in the $x$-direction, with length $L_x$ along $x$ as above and with the cross-sectional ``area" $A$ of the lattice defined such that the total number of unit cells is $A L_x$. The charge density $\nu = Q/(A L_x)$ remains a rational number, and the commutation relation between the flux insertion operator $\hat{D}_x$ and $T_x$ is given by
\beq
\label{eq.commutationTDinhigherD}
\hat{T}_x \hat{D}_x = e^{- 2\pi i A \frac{p}{q}} \hat{D}_x \hat{T}_x \, .
\eeq
The rest of the argument follows in analogy with that in $d=2$ spatial dimensions. Note also that no assumptions were made regarding particle statistics.

\subsection{Dipole moment and translation}

Before generalising the preceding discussion to systems with linear shift symmetries, we observe that the flux insertion operator in Eq.~\eqref{eq.U1LargeGaugeTransformation} is closely related to the operator measuring the dipole moment (or center of mass) in the $x$-direction. In particular, the operator
\beq
\hat{\Tilde{D}}_x = \sum_{\br} x_{\br} \hat{n}_{\br} \, ,
\eeq
measures the dipole moment along $x$. However, $\hat{\Tilde{D}}_x$ is inconsistent with PBC since
\beq
\hat{\Tilde{D}}_x \mapsto \sum_{i} (x_i + L_x) \hat{n}_i = \hat{\Tilde{D}}_x + L_x \hat{Q},
\eeq
under a translation by $L_x$. Thus, the properly defined dipole moment operator for PBC is
\beq
\label{exponentialDipole}
\hat{D}_x = \exp\left(\frac{2\pi i}{L_x} \hat{\Tilde{D}}_x \right) \, ,
\eeq
which exactly matches the U(1) flux insertion operator Eq.~\eqref{eq.U1LargeGaugeTransformation} and whose commutation relation with $\hat{T}_x$ is thus given by Eq.~\eqref{eq.commutationTDinhigherD}. In light of the above, the commutation relation Eq.~\eqref{eq.commutationTDinhigherD} has a different interpretation: eigenstates of a translation invariant system that conserves both charge and dipole moment can carry both translation and dipole moment quantum numbers (in addition to charge) \textit{only} when the charge density is an integer.

\subsection{Constraints on linear shift symmetries}
\label{sec:dipole}

\begin{figure}[t]
\includegraphics[width=0.5\textwidth]{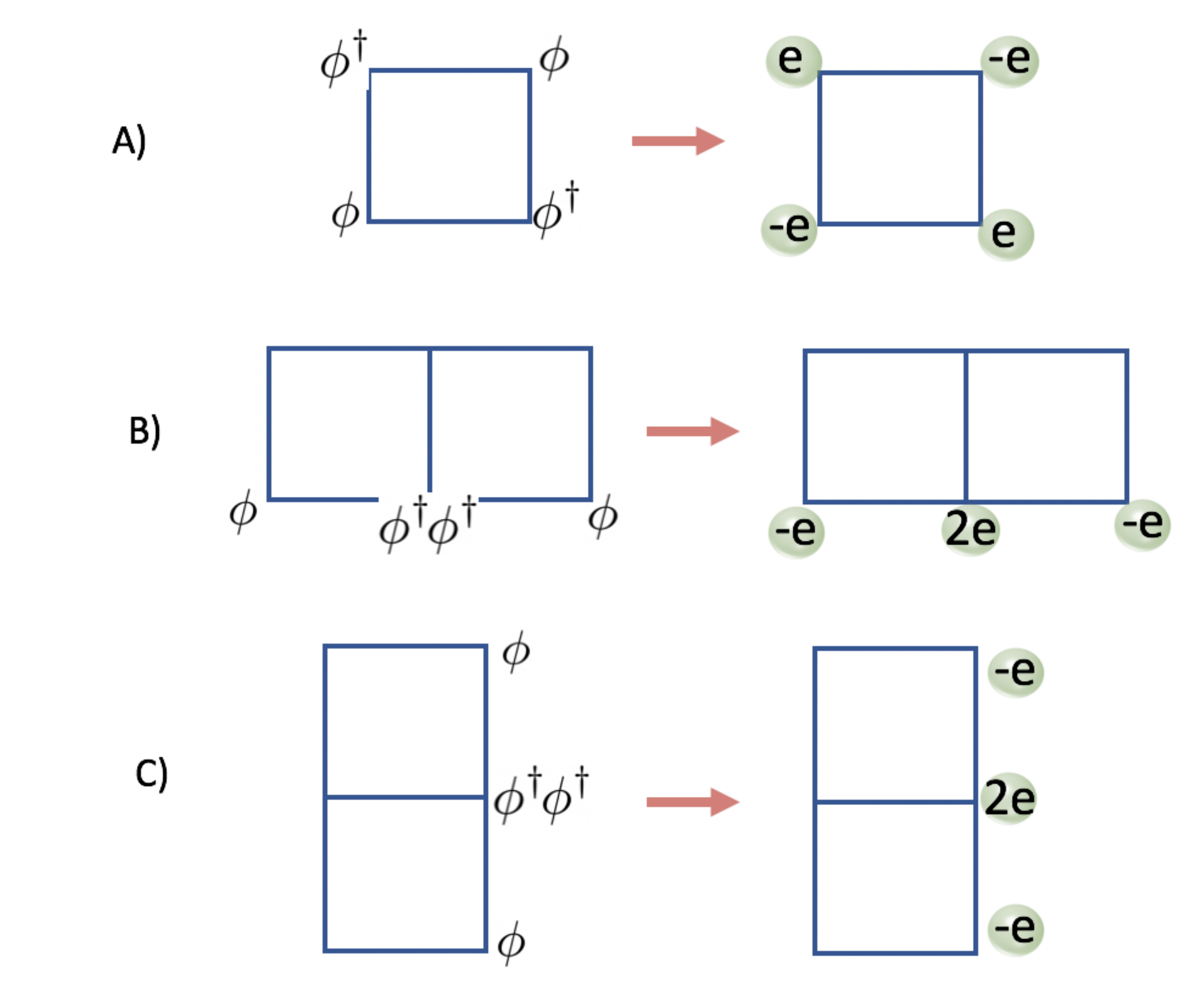}
\caption{Lowest order dipole conserving processes in $d=2$. The motion of isolated charges is forbidden under these, while composite dipolar objects are allowed to hop. For subsytem symmetries, the only allowed process at this order is given by A). } 
\label{fig1}
\end{figure}

Consider an interacting quantum system of bosons on a square lattice such that individual charged bosons\footnote{As before, we assign a fictitious charge $e$ to each particle since we do not require that it carries real electric charge.} do not exhibit any dynamics but inter-bond particle-hole pairs, forming dipoles, are allowed to hop along the $x$ or $y$ direction (see Fig.~\ref{fig1}). The Hamiltonian is
\begin{align} 
\label{dipoleHam}
\hat{H} = &\sum_{\br} \left[\hphi^{\dagger}_{\br} \hphi_{\br + \mathbf{e}_x} \hphi^{\dagger}_{\br + \mathbf{e}_x + \mathbf{e}_y} \hphi_{\br + \mathbf{e}_y} + \hphi^{\dagger}_{\br} \left( \hphi_{\br + \mathbf{e}_x} \right)^2 \hphi^{\dagger}_{\mathbf{r} + 2 \mathbf{e}_x} \right. \nonumber\\
& + \left. \hphi^{\dagger}_{\br}(\hphi_{\br + \mathbf{e}_y})^2 \hphi^{\dagger}_{\mathbf{r} + 2 \mathbf{e}_y}+ \text{h.c} + \dots \right] \, ,
\end{align}
where $\mathbf{e}_i$ is the unit vector in the $i^{th}$-direction. The dots represent higher-order terms which also only allow dynamics for composite, dipolar objects.

Aside from a conventional global U(1) symmetry,
\begin{align}
\label{eq.conventionalU(1)Symmetry}
U(1): \hphi(x,y) \rightarrow e^{i \theta }\hphi(x,y) \, ,
\end{align}
the above Hamiltonian is also invariant under additional U(1) symmetries, referred to as \textit{linear shift} symmetries~\cite{gromov2019}:
\begin{align} 
&U^x(1): \hphi(x,y) \rightarrow e^{i x\theta_x/a }\hphi(x,y)=e^{i x\theta_x }\hphi(x,y) \, , \nonumber\\
&U^y(1): \hphi(x,y) \rightarrow e^{i y\theta_y/a }\hphi(x,y)=e^{i y\theta_y }\hphi(x,y) \, .
\label{sym}
\end{align}
Here, $\theta$, $\theta_x$, and $\theta_y$ are all arbitrary constants, and we have set the lattice spacing $a=1$. Note that $a$ also sets the minimal length scale for the composite dipolar objects, referred to henceforth simply as ``dipoles." While the conventional U(1) symmetry enforces charge conservation, the linear shift symmetries additionally enforce conservation of dipole moment~\cite{fractongauge,gromov2019}. On the lattice, elements of the two U(1) linear shift symmetries are:
\begin{equation}
\begin{split}
\hat{D}_x(\theta_x) & = \exp \left(i \theta_x \sum_{\br} x_{\br} \hat{n}_{\br} \right), \\
\hat{D}_y(\theta_y) & = \exp \left(i \theta_y \sum_{\br} y_{\br} \hat{n}_{\br} \right), \\
\end{split}
\end{equation}
where $x_{\br} (y_{\br})$ refers to the $x (y)$ position of the $\mathbf{r}^{th}$ site. For PBC in both directions, these symmetry operators identically match the flux insertion operators (see Eq.~\eqref{eq.U1LargeGaugeTransformation}) for conventional U(1) symmetries and so also match dipole moment operators along $x$ and $y$ respectively (see Eq.~\eqref{exponentialDipole}). Thus, we will continue using the same notation as before. 

The presence of these unconventional U(1) symmetries forbids the presence of any single boson hopping terms in the many-body Hamiltonian, rendering it a charge insulator whose leading-order dynamics stem from the dipole degrees of freedom. While these symmetries impart a spatially dependent phase shift on the bosonic field $\hphi(x,y)$, they rotate each dipole with a global phase factor; for instance, a dipole operator with a fixed displacement $\mathbf{r} = (r_x,r_y)$ transforms as:
\begin{equation}
\begin{split}
U^x(1): & \quad \hphi(x,y)\hphi^\dagger(x+r_x,y+r_y)  \\
\mapsto & \quad e^{-i r_x \theta_x} \hphi(x,y)\hphi^\dagger(x+r_x,y+r_y)\, , \\
U^y(1): & \quad
\hphi(x,y)\hphi^\dagger(x+r_x,y+r_y) \\
\mapsto & \quad e^{-i r_y \theta_y} \hphi(x,y)\hphi^\dagger(x+r_x,y+r_y) \, .
\end{split}
\end{equation}
Due to the above transformations, we see that the linear shift symmetries act as conventional U(1) symmetries for dipolar objects, thereby generating the conservation of global dipole moment. Nonetheless, these symmetries \textit{cannot} be regarded as internal symmetries, since they are non-trivially intertwined with spatial degrees of freedom (see Ref.~[\onlinecite{gromov2019}] for more details on this point). In other words, the dipole, which consists of at least two \textit{local} particles, is \textit{non-local} in nature. 

Besides imposing conservation of dipole moment, which prohibits the motion of isolated charges, we will make no further assumptions regarding the many-body Hamiltonian. Eq.~\eqref{dipoleHam} can thus be regarded as the most general dipole conserving Hamiltonian, where we have only explicitly written down the lowest-order symmetry allowed terms. Following our discussion of the conventional LSMOH theorem in Sec.~\ref{sec:review}, we now discuss non-trivial constraints imposed on the low energy behavior of $d=2$ translation invariant systems which additionally conserve the global dipole moment. 

We place the system on an $L_x \times L_y$ lattice with PBC in both directions, with both lengths measured in terms of the lattice spacing. Since the phase shifts are spatially dependent, consistency with boundary conditions requires:
\begin{align} 
&\hphi(x=0,y) = e^{i L_x\theta_x}\hphi(x=L_x,y) \, ,\nonumber\\
&\hphi(x,y=0) = e^{i L_y \theta_y}\hphi(x,y=L_y) \, .
\end{align}
Hence, the phase shifts $\theta_x$ and $\theta_y$ can only be integer multiples of $2\pi/L_x$ and $2\pi/L_y$ respectively. In order for the ground state of the system to preserve both linear shift and translation symmetries, $\hat{D}_x(\theta_x)$ and $\hat{D}_y(\theta_y)$ must commute with the translation operators $\hat{T}_x$ and $\hat{T}_y$:
\begin{align} 
\begin{split}
\hat{T}_x \hat{D}_x(\theta_x) \hat{T}_x^{-1} 
=& \exp \left( i \theta_x \sum_{\br} (x_{\br} - 1) \hat{n}_{\br} \right) \\
=& \hat{D}_x(\theta) e^{-i \theta_x L_x L_y \nu } \, , \\
\hat{T}_y \hat{D}_y(\theta_y) \hat{T}_y^{-1}
=& \exp \left(i \theta_y \sum_{\br} (y_{\br} - 1)\hat{n}_{\br} \right)\\
=& \hat{D}_y(\theta_y) e^{-i \theta_y L_x L_y \nu } \, ,
\end{split}
\end{align}
where $\nu = p/q$ is the charge density. Given the constraints imposed on $\theta_x,\theta_y$ by PBC, we see that these operators commute only when $\nu \in \mathbb{Z}$. Consequently, an essential condition for a featureless ground state is an integer charge filling. Since a fractional $\nu$ leads to $q$ degenerate ground states due to the usual LSMOH argument, we henceforth assume that $\nu \in \mathbb{Z}$ in order to uncover the new constraints imposed by dipole moment conservation.

To do this, we can adapt Laughlin's flux-threading argument~\cite{laughlin} in the presence of U(1) linear shift symmetries. Starting in a many-body ground state, we can implement a large gauge transformation for these symmetries by inserting a unit flux quantum $2\pi$ for the dipoles. This procedure leaves the Hamiltonian invariant, and in the presence of a finite excitation gap, the resultant state remains in the ground state manifold. Explicitly, the large gauge transformations (or global flux insertion operators) are given by (see Appendix for details):
\begin{equation}
\label{eq.Dij}
\begin{split}
& \hat{D}_{xx}= \exp \left(\frac{2\pi i}{L_x} \sum_{\br} \frac{x_{\br}^2 \hat{n}_{\br}}{2} \right) , \,  
\hat{D}_{xy}= \exp \left(\frac{2\pi i}{L_x} \sum_{\br} x_{\br} y_{\br} \hat{n}_{\br} \right) , \\
& \hat{D}_{yx} = \exp \left(\frac{2\pi i}{L_y} \sum_{\br} x_{\br} y_{\br} \hat{n}_{\br} \right) , \, 
\hat{D}_{yy} = \exp \left( \frac{2\pi i}{L_y} \sum_{i} \frac{y_{\br}^2 \hat{n}_{\br}}{2} \right) .
\end{split}
\end{equation}
We also require that $L_x$ and $L_y$ be even so that the above operators are consistent with PBC.

Physically, the flux insertion operator $\hat{D}_{ij}$ creates a global flux $2\pi$ along the $i^{th}$-direction for dipoles in the $j^{th}$-direction. These flux insertion operators are in turn related to \textit{quadrupole} operators which measure the many-body quadrupole moments:
\begin{equation}
\label{eq.tildeDij}
\begin{split}
\hat{\Tilde{D}}_{xx} &= \frac{1}{2} \sum_{\br} x_{\br}^2 \hat{n}_{\br}, \,  
\hat{\Tilde{D}}_{yy} = \frac{1}{2} \sum_{\br} y_{\br}^2 \hat{n}_{\br} \, ,    \\
\hat{\Tilde{D}}_{xy} &= \hat{\Tilde{D}}_{yx} = \sum_{\br} x_{\br} y_{\br} \hat{n}_{\br} \, .
\end{split}
\end{equation}
Following the arguments presented in Sec.~\ref{sec:review}, we see that the system hosts a unique, gapped, and symmetry preserving ground state only if the translation operators commute with the large gauge transformations Eq.~\eqref{eq.Dij}. The commutation relations between these operators are:
\begin{align}
&\hat{T}_x \hat{D}_{xx} = \exp \left(-\frac{2\pi i}{L_x} \sum_{\br} x_{\br} \hat{n}_{\br} \right) \hat{D}_{xx} \hat{T}_x = e^{-\frac{2\pi i}{L_x} \hat{\Tilde{D}}_x } \hat{D}_{xx} \hat{T}_x \, , \nonumber\\
&\hat{T}_y \hat{D}_{yy} = \exp\left(-\frac{2\pi i}{L_y} \sum_{\br} y_{\br} \hat{n}_{\br} \right) \hat{D}_{yy} \hat{T}_y = e^{-\frac{2\pi i}{L_y} \hat{\Tilde{D}}_y } \hat{D}_{yy} \hat{T}_y \, , \nonumber\\
&\hat{T}_x \hat{D}_{xy} = \exp\left(-\frac{2\pi i}{L_x} \sum_{\br} y_{\br} \hat{n}_{\br} \right) \hat{D}_{xy} \hat{T}_x =e^{-\frac{2\pi i}{L_x} \hat{\Tilde{D}}_y } \hat{D}_{xy} \hat{T}_x \, , \nonumber\\
&\hat{T}_y \hat{D}_{xy} = \exp \left( -\frac{2\pi i}{L_x} \sum_{\br} x_{\br} \hat{n}_{\br} \right) \hat{D}_{xy} \hat{T}_y = e^{-\frac{2\pi i}{L_x} \hat{\Tilde{D}}_x } \hat{D}_{xy} \hat{T}_y \, , \nonumber\\
&\hat{T}_x \hat{D}_{yx} = \exp \left(-\frac{2\pi i}{L_y} \sum_{\br} y_{\br} \hat{n}_{\br} \right) \hat{D}_{yx} \hat{T}_x = e^{-\frac{2\pi i}{L_y} \hat{\Tilde{D}}_y} \hat{D}_{yx} \hat{T}_x \, , \nonumber\\
&\hat{T}_y \hat{D}_{yx} = \exp \left(-\frac{2\pi i}{L_y} \sum_{\br} x_{\br} \hat{n}_{\br} \right) \hat{D}_{yx} \hat{T}_y= e^{-\frac{2\pi i}{L_y} \hat{\Tilde{D}}_x} \hat{D}_{yx} \hat{T}_y \, .
\label{dip}
\end{align}
Consequently, an interacting bosonic system hosts a unique gapped ground state which preserves both translation and linear shift symmetries (as well as the usual U(1) symmetry), only if: a) the charge filling $\nu \in \mathbb{Z}$, and b) the dipole moments per unit length
\begin{equation*}
\left(\frac{\Tilde{D}_x}{L_x}, \frac{\Tilde{D}_x}{L_y}, \frac{\Tilde{D}_y}{L_x},\frac{\Tilde{D}_y}{L_y}\right) 
\end{equation*}
are also integers.

In order to better understand the implications of the above constraints, we observe that the dipole moment operators $\hat{\tilde{D}}_j$ in Eq.~\eqref{dip} measure the bulk \textit{charge polarization} of the many-body system~\cite{resta1998,oshikawa1,oshikawa2,oshikawasenthil,kang2019}. The many-body polarization $\mathcal{P}_j \, (j=x,y)$ is defined upto an integer shift equivalency~\cite{you2019multipole} \textit{i.e.,} modulo $\mathbb{Z}$ and, in the presence of reflection symmetry, is further quantized to only take values $0,1/2$. Since we have shown that a featureless ground state with conserved dipole moment necessarily has an integer-valued charge polarization, our constraint excludes the possibility that the system is a topological dipole insulator with non-zero net dipole moment~\cite{hughes2019dipole} as that requires either $\mathcal{P}_x = 1/2$ or $\mathcal{P}_y = 1/2$ in the presence of particle-hole or reflection. Thus, we find that the conservation of dipole moment enforces new constraints on the ground state features of interacting lattice systems, with implications for higher order topological insulators.

In the absence of linear shift symmetries, the conventional LSMOH theorem shows that a fractional charge density $\nu = p/q$ implies a degenerate ground state manifold, while a unique symmetric gapped ground state requires integer fillings. However, translation invariant systems which additionally conserve dipole moment can have a degenerate ground state manifold with a finite excitation gap even for integer charge fillings. To see this, let us consider $\nu \in \mathbb{Z}$ but fractional charge polarization (equivalently, fractional dipole moment per length). Then, Eq.~\eqref{dip} dictates that the ground state degeneracy (GSD) is at least
\beq
\begin{split}
\mathrm{GSD} & \geq \mathrm{lcm}\left( \frac{L_x}{\mathrm{gcd}(L_x,\Tilde{D}_x)},
\frac{L_x}{\mathrm{gcd}(L_x,\Tilde{D}_y)},
\frac{L_y}{\mathrm{gcd}(L_y,\Tilde{D}_y)}
\right) \\
&\times \mathrm{lcm}\left( \frac{L_y}{\mathrm{gcd}(L_y,\Tilde{D}_y)},
\frac{L_x}{\mathrm{gcd}(L_x,\Tilde{D}_x)},
\frac{L_y}{\mathrm{gcd}(L_y,\Tilde{D}_x)}
\right),
\end{split}
\eeq
where ``$\mathrm{lcm}$" refers to the least common multiple and the ``$\mathrm{gcd}$" refers to the greatest common divisor. When the charge polarization is constant over the lattice, \textit{i.e.,}
\beq
\nu_x = \frac{\Tilde{D}_x}{L_xL_y} = \frac{p_x}{q_x},\quad
\nu_y = \frac{\Tilde{D}_y}{L_xL_y} = \frac{p_y}{q_y},
\eeq
with $p_x$ and $q_x$ coprime and $p_y$ and $q_y$ coprime, the above simplifies to
\beq
\mathrm{GSD} \geq \mathrm{lcm}\left( q_x,q_y \right)^2 \, ,
\eeq
where we have further assumed that $L_x$ and $L_y$ are coprime with $q_x$ and $q_y$. Similarly to our discussion in Sec.~\ref{sec:review}, this assumption is justified in the thermodynamic limit. We also note that while we have discussed interacting bosonic systems here, the above argument readily generalizes to systems of interacting particles with arbitrary statistics.

\subsection{Generalization to polynomial shift symmetries}
\label{sec:general}

So far, we have only considered dipole moment conserving systems. In principle, one can also consider systems which conserve even higher moments, such as quadrupole or octupole. A systematic study of these systems was undertaken by Gromov~\cite{gromov2019}, who illustrated that higher moment conservation laws naturally arise in systems with U(1) polynomial shift symmetries. Specifically, when the independent symmetry generators are chosen such that they include all monomials up to some fixed order $m$, then multipoles of up to order $m$ will be conserved. Large gauge transformations of these symmetries can be derived and are given by multipoles of order $m+1$; one can then derive the commutation relations between these operators and the translation operators $\hat{T}_j$. The phase factor which appears in these commutation relations is a function of the system size as well as of multipoles of order $\leq m$. Finally, one can derive the constraints placed on the ``densities" of composite multipolar objects under the assumption of a unique gapped ground state which preserves both translation and all U(1) polynomial shift symmetries up to order $m$.

The above procedure has been implemented for $d=2$ systems invariant under arbitrary polynomial symmetries, with details presented in the Appendix so as not to clutter the main text with technical details and also since we do not find any results which are qualitatively distinct from those described in the preceding section.


\section{Constraints on U(1) Subsystem Symmetries}
\label{sec:subsystem}

A subsystem symmetry in $d$ spatial dimensions consists of independent symmetry operations acting on a set of $D$-dimensional subsystems, with $0<D<d$. In $d=2$, these can be lines or fractals, while in $d=3$, the subsystems can be lines, planes, or fractals. Analogously to ordinary global symmetries, subsystem symmetries can be spontaneously broken~\cite{xumoore} or can protect non-trivial topological phases, dubbed subsystem protected topological (SSPT) phases~\cite{yizhi1,yizhi2} which host gapless boundary modes. In this section, we first establish constraints imposed by the presence of both U(1) subsystem and translation symmetries in $d=2$, after which we discuss implications of such LSM-type constraints for SSPT phases.

\subsection{Constraints from subsystem and translation symmetry}

Consider a system of interacting charged bosons on an $L_x \times L_y$ square lattice invariant under line-like U(1) subsystem symmetries, so that the total charge on each row and column is separately conserved. The net charge on each row and column is given respectively by
\begin{equation}
\begin{split}
\hat{Q}_{x,j} &= \sum_{\mathbf{r} \, \in \, j^{th}\text{ row}} \hat{n}_{\br}, 
\quad \forall \; j=1,2,\ldots,L_y.
\\
\hat{Q}_{y,j} &= \sum_{\mathbf{r} \, \in \, j^{th}\text{ col}} \hat{n}_{\br},
\quad \forall \; j=1,2,\ldots,L_x.   \\
\end{split}
\end{equation}
Group elements of the line-like U(1) symmetries are given by
\begin{equation}
\begin{split}
\hat{U}_{x,j}(\theta) = e^{i \theta \hat{Q}_{x,j}}, \;\;
\hat{U}_{y,j}(\theta) = e^{i \theta \hat{Q}_{y,j}} \, ,
\end{split}
\end{equation} 
respectively, where $\theta$ is an arbitrary constant. A generic Hamiltonian for an interacting system respecting such symmetries is given by
\begin{align} 
\hat{H} = \sum_{\br} \left[ \hphi^{\dagger}_{\br}\hphi_{\br + \mathbf{e}_x}\hphi^{\dagger}_{\br + \mathbf{e}_x + \mathbf{e}_y}\hphi_{\br + \mathbf{e}_y} + \text{h.c} + \dots \right] \, ,
\end{align}
where only the lowest order symmetry-preserving term is written explicitly and the dots represent higher-order symmetry respecting terms. In contrast with dipole conserving systems (see Sec.~\ref{sec:dipole}), which allow dipoles to hop freely while constraining charge motion, subsystem symmetries additionally constrain the motion of dipoles, as depicted in Fig.~\ref{fig1}. In particular, a dipole is only allowed to hop in the direction perpendicular to its orientation. 

Suppose we now impose PBC in both the $x$ and $y$-directions. In order for the system to preserve both translation and U(1) subsystem symmetries, the total charge in each row and column must be uniform \textit{i.e.,}
\begin{equation}\label{eq.uniformCharge}
\begin{split}
\hat{Q}_{x,j} = \hat{Q}_{x,j+1}, \quad\forall\; j=1,2,\ldots,L_y,    \\
\hat{Q}_{y,j} = \hat{Q}_{y,j+1}, \quad\forall\; j=1,2,\ldots,L_x.    \\ 
\end{split}
\end{equation} 
Starting from the many-body system, we can further implement a large gauge transformation by adiabatically threading a $2\pi$ flux for the charge on a specific row or column, which is implemented by the operators
\begin{equation} 
\begin{split}
\hat{U}_{x,j} &= \exp\left(\frac{2\pi i}{L_x} \sum_{\mathbf{r} \, \in \, j^{th}\text{ row}} x_{\br} \hat{n}_{\br} \right), \forall\;
j=1,\ldots,L_y \, ,\\
\hat{U}_{y,j} &= \exp\left(\frac{2\pi i}{L_y} \sum_{\mathbf{r} \, \in \, j^{th}\text{ col}} y_{\br} \hat{n}_{\br} \right), \forall\;
j=1,\ldots,L_x \, , \\
\end{split}
\end{equation}
whose commutation relations with the translation operators $\hat{T}_x$ and $\hat{T}_y$ are
\begin{equation}
\label{eq.subsystemLSM}
\begin{split}
\hat{T}_{x} \hat{U}_{x,j} \hat{T}_{x}^{-1} &= \hat{U}_{x,j} \exp\left(- \frac{2\pi i}{L_x} \hat{Q}_{x,j} \right), \\
\hat{T}_{y} \hat{U}_{x,j} \hat{T}_{y}^{-1} &= \hat{U}_{x,j+1}, \quad \forall \; j=1,2,\ldots,L_y, \\
\hat{T}_{y} \hat{U}_{y,j} T_{y}^{-1} &= \hat{U}_{y,ij} \exp\left(- \frac{2\pi i}{L_y} \hat{Q}_{y,j} \right),    \\
\hat{T}_{x} \hat{U}_{y,j} \hat{T}_{x}^{-1} &= \hat{U}_{y,j+1}, \quad \forall \; j=1,2,\ldots,L_x \, . \\
\end{split}
\end{equation}
In the presence of a finite excitation gap, the adiabatic theorem guarantees that the state obtained after the flux-threading procedure remains in the ground state manifold. Following the arguments presented in Sec.~\ref{sec:multipole}, we then see that in order to obtain a unique gapped ground state which preserves all symmetries, $\hat{T}_x$ and $\hat{T}_y$ must commute with $\hat{U}_{x,j}$ and $\hat{U}_{y,j}$. In turn, this implies that the uniform charge densities on each row and column
\beq
\nu_x  = \frac{Q_{x,j}}{L_x}, \quad \nu_y = \frac{Q_{y,j}}{L_y}  \, ,
\eeq
are restricted to take integer values, and the dipole moments in each row and column must be uniform
\begin{equation}\label{eq.uniformDipole}
\begin{split}
\sum_{\mathbf{r} \, \in \,  j^{th}\text{ row}} x_{\br} \hat{n}_{\br} &= \sum_{\mathbf{r} \in (j+1)^{th}\text{ row}} x_{\br} \hat{n}_{\br}, \quad \forall\;
j=1,\ldots,L_y.\\
\sum_{\mathbf{r} \in j^{th}\text{ col}} y_{\br} \hat{n}_{\br} &= \sum_{\mathbf{r} \in (j+1)^{th}\text{ col}} y_{\br} \hat{n}_{\br}, \quad \forall\;
j=1,\ldots,L_x. \\
\end{split}
\end{equation}
Thus, the above argument eliminates the possibility of a unique, symmetry-preserving gapped ground state if either the net charge in each row/column is non-uniform or if the net charge in each row/column is uniform but has fractional filling $\nu_j \notin \mathbb{Z} \, (j=x,y)$.

Let us consider the latter scenario first, where the total charge and dipole moment per row/column are uniform (so Eqs.~\eqref{eq.uniformCharge} and~\eqref{eq.uniformDipole} are satisfied) while the charge densities are fractional: $\nu_x = p_x/q_x$ and $\nu_y = p_y/q_y$. Then, we find that
\begin{equation}
Q_{x,i} L_y = Q_{y,j} L_x \Longleftrightarrow
\nu_x = \frac{Q_{x,i}}{L_x} = \frac{Q_{y,j}}{L_y} = \nu_y \, ,
\end{equation}
and secondly that the ground state manifold is necessarily degenerate, with the degeneracy at least
\begin{equation}
\mathrm{GSD} \geq q_x q_y = q_x^2 \, ,
\end{equation}
provided that $q_x (q_y)$ and $p_x (p_y)$ are coprime. On the other hand, if the net charge per row/column is not uniform, the GSD generated by the algebra in Eq.~\eqref{eq.subsystemLSM} is upper bounded by $L_x L_y$ because the maximally commuting set of operators only contains $\hat{T}_x$ and $\hat{T}_y$. Similarly, in $d \geq 3$ dimensions, the GSD generated by the above argument is upper bounded by the volume of the system. This volume law degeneracy corresponds to a trivial integrable model with all sites decoupled.

To enhance the precise lower-bound on the GSD predicted by the above argument, we need to introduce additional ingredients besides the translation and flux insertion operators considered above. A natural extension is to introduce fractional charges into the system: in Ref.~[\onlinecite{oshikawasenthil}], it was shown that the existence of excitations carrying fractional charge $p/q$ (in units of $e$) in a $d=2$ gapped system indicates a $q$-fold degenerate ground state manifold on the two-torus. The additional operator introduced in this setting corresponds to one that drags a fractionalized excitation around a non-trivial cycle of the torus. This argument can be straightforwardly adapted to the case of U(1) subsystem symmetries. Specifically, we find that line-like subsystem symmetries in both $d=2,3$ need not give rise to a non-trivial GSD; however, planar subsystem symmetries in a $d=3$ gapped, translation invariant system with fractionalized charges $p/q$ imply a GSD which grows exponentially with the linear extent of the system \textit{i.e.,} GSD $\sim q^L$. A detailed derivation of this statement is presented in the Appendix.

\subsection{Implications for SSPT phases}
\label{sec:implications}

We now briefly comment on applications of the LSM-type constraints explored here to the classification and characterization of interacting SSPT phases~\cite{yizhi1,yizhi2,devakulfractal,strongsspt,shirleygauging,shirleytwisted} via boundary anomalies. First, we recall the relation between LSM-type constraints and the surface anomalies of SPT phases~\cite{cheng2016,hsieh2016,cho2017,huang2017,jian2018,metlitski2018,thorngren2018,cheng2019fermionic,jiang2019generalized,else2019topological,yang2018,bultinck2018}, which has been established through the bulk-boundary correspondence of crystalline SPT phases.

The generalized LSM theorem states that a $d$-dimensional interacting quantum system invariant under an internal symmetry $G$ and some spatial symmetry $S$ (\textit{e.g.,} lattice translation $\mathbb{Z}^d$) cannot support a featureless gapped phase with a unique ground state invariant under $G\times S$. In parallel with this no-go theorem, an SPT boundary which is anomalous under $G$ and an on-site symmetry $\tilde{S}$ cannot host a featureless gapped surface state without breaking $G \times \tilde{S}$. The low energy effective theory of the surface state of a $(d+1)$-dimensional SPT with $G\times\tilde{S}$ is thus found to be equivalent to that of a $d$-dimensional lattice model with symmetry group $G \times S$, where the lattice symmetry $S$ can be interpreted as an internal symmetry $\tilde{S}$ under ultraviolet regularization. This mapping has led to fruitful results: since an SPT boundary with $G \times S$ symmetry does not permit a trivial gapped boundary, the corresponding lattice spin system cannot support a featureless gapped phase. Analogously, if a generalized LSM theorem implies the absence of a featureless gapped ground state, one can extrapolate from this dual mapping that the corresponding SPT surface theory is anomalous and hence relies on a non-trivial bulk state.

Following the above perspective, one can propose a general criterion for the existence of some specific SSPT phase by relating its boundary anomaly to an LSM constraint. Specifically, consider a $d$-dimensional SSPT state protected by both a subsystem symmetry $G^{sub}$ and an on-site global symmetry $S$, whose symmetric boundary theory can be mapped onto a lower-dimensional lattice model invariant under $G^{sub}$ and a lattice symmetry whose action is similar to that of $S$ at the SSPT boundary. This mapping between an on-site symmetry at the boundary of a $d$-dimensional SSPT and a crystalline symmetry in a $(d-1)$-dimensional lattice model allows us to restrict the conditions under which an SSPT may host a featureless gapped boundary. In particular, the absence of a featureless gapped phase invariant under $G^{sub}$ and some lattice symmetry $\tilde{S}$ in 2d implies that a 3d SSPT phase does not admit a trivial gapped boundary in the presence of $G^{sub}$ and an on-site symmetry $S$, which plays the role of $\tilde{S}$ at the boundary. Consequently, the ungappable surface state is anomalous and must be accompanied by a non-trivial bulk SSPT. Hence, the constraints developed in the previous section for systems with line-like U(1) subsystem symmetries and lattice translation symmetry imply constraints on the surface states of generic interacting SSPT phases~\cite{you2019higher} in $d=3$. 


\section{Outlook}
\label{sec:cncls}

In this work, we have derived LSMOH constraints for lattice translation invariant systems with generalized U(1) symmetries and discussed their implications for the existence of featureless gapped phases with unique symmetry preserving ground states. We expect that the ideas developed here can be further generalized to subsystem symmetries besides U(1) as well as to space group symmetries beyond lattice translation, leading towards a more complete theory of LSM-type constraints for systems with emergent fractonic behavior. Further research in this direction is also likely to shed light on symmetry enriched fracton phases~\cite{yizhi2}, which have yet to be studied in much detail and which may potentially display fascinating phenomenology. Moreover, conventional LSM-type arguments have found applications in the study of symmetry enriched quantum spin liquids as well as deconfined quantum critical points; we hence anticipate that our results will aid in the search for exotic quantum spin liquids with emergent fractonic behavior as well as subsystem invariant quantum criticality.


\begin{acknowledgements}

We are grateful to Andrei Bernevig, Sheng-Jie Huang, Zhu-Xi Luo, Sid Parameswaran, and Michael Pretko for stimulating discussions. Y. Y. and A. P. are supported by fellowships at the PCTS at Princeton University. This work was performed in part at the Aspen Center for Physics, which is supported by National Science Foundation grant PHY-1607611.

\end{acknowledgements}


\newpage 

\bibliography{library}


\renewcommand{\bibnumfmt}[1]{[S#1]}
\renewcommand{\theequation}{S\arabic{equation}}
\renewcommand{\thefigure}{S\arabic{figure}}
\renewcommand{\thetable}{S.\Roman{table}}

\setcounter{section}{0}
\setcounter{equation}{0}
\setcounter{figure}{0}
\setcounter{table}{0}

\clearpage
\onecolumngrid


\section*{Appendix}
\label{appendix}

\onecolumngrid

\subsection{Large gauge transformations for U(1) linear shift symmetries}
\label{app:A}

In this Appendix, we systematically derive the large gauge transformations, or flux insertion operators, for global and linear shift U(1) symmetries. For simplicity, we restrict our discussion to the continuum, with generalizations to arbitrary dimensions and lattices straightforward. Periodic boundary conditions (PBC) in all directions are assumed.

\subsubsection{Conventional U(1) symmetry} 

In this section, we derive the flux insertion operators for a conventional U(1) global symmetry in two dimensions. The U(1) transformation for a charged field is given by Eq.~\eqref{eq.conventionalU(1)Symmetry} in the main text. Without loss of generality, we only discuss bosonic fields here. The kinetic energy, which is symmetric under the U(1) global symmetry, is given by
\begin{equation}
\hat{H}_{k} = \frac{1}{2} \partial_j \hphi^\dagger \partial_j \hphi \, ,
\end{equation}
where the repeated index $j$ sums over $x$ and $y$. In order to find the flux insertion operators, we first gauge the global U(1) symmetry. By the usual minimal coupling procedure, the kinetic energy gets modified as
\begin{equation}
\hat{H}_{k} \mapsto \frac{1}{2} 
\left[\left( i \partial_j + \hat{a}_j \right)\hphi\right]^\dagger
\left[\left( i \partial_j + \hat{a}_j \right)\hphi\right] \, ,
\end{equation}
which is invariant under the following gauge transformations:
\begin{equation}
\begin{split}
\hphi(x,y) &\mapsto e^{i \theta(x,y)} \hphi(x,y), \\
\hat{a}_{j}(x,y) &\mapsto \hat{a}_{j} (x,y) + \partial_{j} \theta(x,y), \quad j =x,y. \\
\end{split}
\end{equation}
where $\theta(x,y)$ is an arbitrary function of the coordinates. The variable conjugate to $\hat{a}_j(x,y)$ is the electric field operator $\hat{e}_j(x,y)$, with the canonical commutation relations
\begin{equation}
\lbrack \hat{a}_j(x,y), \hat{e}_k (x^\prime,y^\prime)\rbrack = i \delta_{jk} \delta(x-x^\prime)\delta(y-y^\prime) \, .
\end{equation}
The Gauss' law constraint is given by
\begin{equation}
\partial_j \hat{e}_j(x,y) = \hat{n}(x,y) \, ,
\end{equation}
where $\hat{n}(x,y)$ is the density operator in the continuum.

The flux encircled by a closed loop $\gamma$ is measured by the Wilson operator:
\begin{equation}
\hat{W}\left( \gamma \right) = \oint_{\gamma} \hat{a}_j dx^j \, .
\end{equation}
In particular, the fluxes through the holes of the torus are measured by the operators
\begin{equation}
\hat{W}(C_x) = \oint_{C_x} \hat{a}_j dx^j, \quad
\hat{W}(C_y) = \oint_{C_y} \hat{a}_j dy^j,
\end{equation}
where $C_x$ and $C_y$ are the non-trivial cycles in the two directions. To find the flux insertion operator, a useful observation is that the operator
\begin{equation}
\begin{split}
\hat{D}(f(x,y)) =& \exp \left(i \iint f(x,y) \hat{n}(x,y) dxdy \right) \, ,
\end{split}
\end{equation}
with $f(x,y)$ arbitrary for now, has the following commutation relation with $\hat{W}(C_x)$:
\begin{equation}
\begin{split}
\hat{D}(f) \hat{W}(C_x) \left( \hat{D}(f) \right)^{-1}
=& \hat{W}(C_x) + \lbrack i \iint f(x_1,y_1) \hat{n}(x_1,y_1) dx_1dy_1, \oint_{C_x} \hat{a}_x(x_2,y_2) dx_2 \rbrack    \\
=& \hat{W}(C_x) + \lbrack i \iint f \partial_j \hat{e}_j dxdy, \oint_{C_x} \hat{a}_x dx \rbrack    \\
=& \hat{W}(C_x) +i \iint dx_1 dy_1 \oint_{C_x} dx_2 \lbrack f(x_1,y_1) \partial_j \hat{e}_j(x_1,y_1), \hat{a}_x(x_2,y_2)\rbrack \\
=& \hat{W}(C_x) + \iint dx_1dy_1 \oint_{C_x} dx_2 f(x_1,y_1) \partial_{x_1}  \delta(x_1-x_2)\delta(y_1-y_2)\\
=& \hat{W}(C_x) - f(0,y_2) + f(L_x,y_2) \, ,
\end{split} 
\end{equation}
where in the first equality, the Baker-Campbell-Hausdorff equality is applied. Therefore, we see that in order for $\hat{D}(f)$ to be the operator which corresponds to threading a unit quantum flux $2\pi$, as measured by $\hat{W}(C_x)$, we necessarily require that
\begin{equation}
- f(0,y_2) + f(L_x,y_2) = 2\pi, \quad \forall\; y_2,
\end{equation}
Hence, without loss of generality, we can choose
\begin{equation}
f(x,y) = \frac{2\pi}{L_x} x.
\end{equation}
Therefore, we obtain the continuum limit of the flux insertion operator Eq.~\eqref{eq.U1LargeGaugeTransformation} used in the main text. Note that the large gauge transformation is by no means unique. Any large gauge transformation followed by another ``small" gauge transformation is still a large gauge transformation.

\subsubsection{Linear shift symmetry and flux insertion operators}
\label{sec.QuadraticShiftSymmetryAndFluxOperators}

In this section, we derive the flux insertion operators for the U(1) linear shift symmetry. To demonstrate the generality of this approach, we work in $d=3$ spatial dimensions here, with the $d=2$ case following immediately. The U(1) linear shift symmetry for a charged bosonic field is given by Eq.~\eqref{sym} in the main text. The kinetic energy which is symmetric under this symmetry is
\begin{equation}
\hat{H}_{k} = \frac{1}{8}m 
\partial_{i} \partial_{j} \hat{\varphi}
\partial_{i} \partial_{j} \hat{\varphi} \, ,
\end{equation}
where the repeated indices $i,j$ imply summation over $x$, $y$, and $z$. Importantly, the field $\hat{\varphi}$ represents only the phase of $\hphi$.

Following the same logic as in the preceding section, we first gauge the U(1) linear shift symmetry. The minimal coupling gives rise to~\cite{fractongauge}
\begin{equation}
\hat{H}_{k} = \frac{1}{8}m 
\left[\left(\partial_{i} \partial_{j} - \hat{a}_{ij} \right) \hat{\varphi}\right]
\left[\left( \partial_{i} \partial_{j} - \hat{a}_{ij} \right)\hat{\varphi} \right] \, ,
\end{equation}
which is invariant under the gauge transformation:
\begin{equation}\label{eq.symmetricGaugeSymmetry}
\begin{split}
\hat{\varphi}(x,y,z) &\mapsto \hat{\varphi}(x,y,z) + \theta(x,y,z), \\
\hat{a}_{ij}(x,y,z) &\mapsto \hat{a}_{ij}(x,y,z) + \partial_{i} \partial_{j} \theta(x,y,z),
\end{split}
\end{equation}
where $\theta(x,y,z)$ is any function of the coordinates. The variables conjugate to the gauge fields $\hat{a}_{ij}$ are the generalized electric fields $\hat{e}_{ij}$. Their commutation relations are:
\begin{equation}
\lbrack
\hat{e}_{jk}(x,y,z),
\hat{a}_{mn}(x^\prime,y^\prime,z^\prime)
\rbrack
= i \delta_{jm}\delta_{kn}
\delta(x-x^\prime)
\delta(y-y^\prime)
\delta(z-z^\prime) \, ,
\end{equation}
with the following Gauss' law constraint
\begin{equation}
\partial_{i}\partial_{j} \hat{e}_{ij} = \hat{n} \, ,
\end{equation}
where $\hat{n}$ measures the charge density. 

The flux encircled by a closed path $\gamma$ is measured by Wilson operators:
\begin{equation}
\hat{W}_i(\gamma) = \oint_{\gamma} \hat{a}_{ij} dx^j \, .
\end{equation}
Note that this operator is gauge invariant under the gauge transformations given in Eq.~\eqref{eq.symmetricGaugeSymmetry}. When $\gamma$ is chosen to be one of the non-trivial cycles of the 3-torus, the flux measured by the Wilson operators is inserted through the holes of the 3-torus, \textit{i.e.,} 
\begin{equation}
\hat{W}_i(C_{j}) = \int_{C_{j}} \hat{a}_{ij} dx^j, \quad \forall\; i,j=x,y,z.
\end{equation}
To construct the flux insertion operators, we will look at the commutation relations between $\hat{W}_i(C_{j})$ and the operator
\begin{equation}
\label{eq.qradraticShiftSymmetryFluxAnsatz}
\hat{D}(f) = \exp \left(i \iiint f(x,y,z) \hat{n}(x,y,z) dx dy dz \right) \, ,
\end{equation}
with $f$ any function of $x$, $y$, and $z$ for now. The commutation relation between $\hat{D}(f)$ and $\hat{W}_x(C_{x})$ is:
\begin{equation}
\label{eq.qradraticShiftSymmetryFluxAnsatzVerify}
\begin{split}
\hat{D}\left( f \right) \hat{W}_x(C_{x}) \hat{D}\left( f \right)^\dagger =& \hat{W}_x(C_{x}) + \lbrack
i \iiint f(x_1,y_1,z_1) \hat{n}(x_1,y_1,z_1) dx_1 dy_1 dz_1,
\int_{C_{x}} \hat{a}_{xx}(x_2,y_2,z_2) dx_2
\rbrack \\
=& \hat{W}_x(C_{x}) + \lbrack
i \iiint f(x_1,y_1,z_1) \partial_{m} \partial_{n} \hat{e}_{mn}(x_1,y_1,z_1) dx_1 dy_1 dz_1,
\int_{C_{x}} \hat{a}_{xx}(x_2,y_2,z_2) dx_2
\rbrack \\
=& \hat{W}_x(C_{x})
- \iiint dx_1 dy_1 dz_1
\int_{C_{x}} dx_2
f(x_1,y_1,z_1) \partial_{x_1}^2
\delta(x_1-x_2) \delta(y_1-y_2) \delta(z_1-z_2) \\
=& \hat{W}_x(C_{x})
- \int dx_1
\int_{C_{x}} dx_2
\partial_{x_1}^2 f(x_1,y_2,z_2)    \\
=& \hat{W}_x(C_{x})
- \int dx_1
\partial_{x_1}^2 f(x_1,y_2,z_2)    \\
=& \hat{W}_x(C_{x})
+ \partial_{x} f(0,y_2,z_2) - \partial_{x} f(L_x,y_2,z_2)  \, .
\end{split}
\end{equation}
In order for $\hat{D}(f)$ to insert a $2\pi$ flux for $\hat{W}_{x}(C_x)$, we choose 
\begin{equation}
\label{eq.ansatzForD(f)}
f(x,y,z)= \frac{\pi}{L_x} x^2 \, .
\end{equation}
Thus, we have obtained the continuum version of the first operator in Eq.~\eqref{eq.Dij} (see main text). The remaining operators in Eq.~\eqref{eq.Dij} can be analogously constructed. Note again that Eq.~\eqref{eq.ansatzForD(f)} is by no means unique.


\subsection{Polynomial Shift Symmetry and Multipole Operators}
\label{app:B}

In this Appendix, we generalize the results obtained for linear shift symmetries (obtained in the main text and in the previous section) to arbitrary polynomial shift symmetries in two dimensions. Requiring a unique gapped ground state which preserves translation as well as all polynomial shift symmetries up to a given order will constrain the ``densities" of multipoles. We consider an $L_x \times L_y$ lattice, with periodic boundary conditions in the $x$-direction. Boundary conditions in the $y$-direction will be separately specified. 

\subsubsection{Polynomial Shift Symmetries and Translation Symmetry}

To begin with, we note that a charged bosonic field transforms under a polynomial shift symmetry as~\cite{gromov2019}: 
\begin{equation}
\label{eq.generalPolynomial}
\hphi(x,y) \rightarrow \exp\left( i \sum_{a\ge 0, b\ge 0, m \ge a+b} \theta_{ab} x^{a}y^{b} \right) \hphi(x,y) \, ,
\end{equation}
where $\theta_{ab}$ are all independent constants. The phase factors contain all monomials up to some fixed order $m$, such that there exist $(m+1)(m+2)/2$ independent symmetry generators:
\begin{equation}
\hat{D}_{ab}(\theta_{ab}) = \exp \left( i \theta_{ab} \sum_{\br} x_{\br}^a y_{\br}^b \hat{n}_{\br} \right), \quad
\forall\;
a\ge 0, b\ge 0, m \ge a+b \, .
\end{equation}
where the summation is taken over all the lattice sites $\br$. As first pointed out in Ref.~[\onlinecite{gromov2019}], this unconventional symmetry underlies the conservation multipole moments up to order $m$ in a system with a global U(1) symmetry.

Suppose we now place the system on a cylinder \textit{i.e.,} open boundary conditions in the $y$-direction. The polynomial shift symmetry is compatible with PBC in the $x$-direction when:
\begin{equation}
\theta_{ab} ((x+L_x)^a-x^a) y^b \in 2\pi \mathbb{Z},
\quad\forall\;
x=0,1,\ldots,L_x-1; \,
y=0,1,\ldots,L_y-1;
\quad\forall\;
a\ge 0, b\ge 0, m \ge a+ b \, .
\end{equation}
Therefore, the constants $\theta_{ab}$ are quantized as follows:
\begin{equation}\label{eq.generalPolynomialQuantizationOfTheta}
\begin{split}
\theta_{ab} \in \frac{2\pi}{L_x^a}\mathbb{Z},\quad
\forall\;
a\ge 1, b\ge 0, m \ge a+ b,
\end{split}
\end{equation}
while $\theta_{0b}$ are not constrained. In order for the ground state manifold to preserve both polynomial shift as well as lattice translation symmetry, the polynomial shift symmetry generators $D_{ab}$ must commute with the translation operator $T_x$. This requires that
\begin{equation}
\theta_{ab} \sum_{\br} \left((x_{\br} + 1)^a - x_{\br}^a \right) y_{\br}^b \hat{n}_{\br}
=
\theta_{ab} \sum_{\br}
\sum_{s=0}^{a-1} \combinationNumber{a}{s} x_{\br}^s y_{\br}^b \hat{n}_{\br}
\in 2\pi \mathbb{Z}, \quad
\forall\;
a\ge 0, b\ge 0, m \ge a+ b \, , 
\end{equation}
which simplifies once we account for the quantization of $\theta_{ab}$ in Eq.~\eqref{eq.generalPolynomialQuantizationOfTheta}:
\begin{equation}
\label{eq.generalPolynomialConstraint1}
\begin{split}
&\frac{2\pi}{L_{x}^{a}} \sum_{\br}
\sum_{s=0}^{a-1} \combinationNumber{a}{s} x_{\br}^s y_{\br}^b \hat{n}_{\br} \in 2\pi \mathbb{Z}, 
\quad\forall\;
a\ge 1, b\ge 0, m \ge a+ b \, ,
\end{split}
\end{equation}
which constrains the multipoles.

\subsubsection{Large Gauge Transformation}

In order to derive the large gauge transformations, we now place the system on the two-torus (PBC in both directions). The kinetic energy which is symmetric under the polynomial shift symmetry is
\begin{equation}
H_{k} = \sum_{l=0}^{m+1} \left[ c_l \left(
\partial_{x}^{l}\partial_{y}^{m+1-l} \hat{\varphi} \right) \left( \partial_{x}^{l}\partial_{y}^{m+1-l} \hat{\varphi} \right) \right] \, ,
\end{equation}
where $\hat{\varphi}$ represents the phase of the bosonic field $\hphi$ and where the $c_l$ are constant parameters. Following standard procedure, we gauge the polynomial shift symmetry so that kinetic energy minimally couples to the gauge fields: 
\begin{equation}
H_{k} \mapsto \sum_{l=0}^{m+1} \left[c_l \left[
\left( 
\partial_{x}^{l}\partial_{y}^{m+1-l} 
- \hat{a}_{l}
\right) \hat{\varphi} \right] \left[
\left(
\partial_{x}^{l}\partial_{y}^{m+1-l} 
- \hat{a}_{l}
\right)\hat{\varphi} \right] \right] \, ,
\end{equation}
invariant under the gauge transformation:
\begin{equation}
\label{eq.generalPolynomialGaugeTransformation}
\begin{split}
\hat{\varphi} \, \mapsto & \, \hat{\varphi} + \theta,   \\
\hat{a}_{m} \, \mapsto & \, \hat{a}_{m} + \partial_{x}^{m}\partial_{y}^{n+1-m} \theta, \quad\forall\;
m=0,1,\ldots,n+1.
\end{split}
\end{equation}
where $\theta$ can be any function of the coordinates. The conjugate variables of the gauge fields are generalized electric fields, with canonical commutation relations:
\begin{equation}
\lbrack
\hat{e}_{l}(x,y), \hat{a}_{l^\prime}(x^\prime,y^\prime) 
\rbrack
= i \delta_{ll^\prime}\delta(x-x^\prime)\delta(y-y^\prime).
\end{equation}
The Gauss' law constraint is also enforced:
\begin{equation}
\sum_{l} \partial_{x}^{l}\partial_{y}^{m+1-l} \hat{e}_{l}(x,y) = \hat{n}(x,y) \,,
\end{equation}
where the right hand side denotes the charge density operator.

The flux encircled by a closed path $\gamma$ is measured by the operator:
\begin{equation}
\hat{W}_l(\gamma) = \oint_{\gamma} \hat{a}_{l} dr, \quad \forall\; l=0,1,\ldots,m+1.
\end{equation}
Note that this operator is gauge invariant under the small gauge transformation in Eq.~\eqref{eq.generalPolynomialGaugeTransformation} and that when the closed path $\gamma$ is a non-trivial cycle of the torus, the flux measured by this operator is threaded through one of the holes of the torus.

Following the same methodology as in the previous section, which led to Eqs.~\eqref{eq.qradraticShiftSymmetryFluxAnsatz} and~\eqref{eq.qradraticShiftSymmetryFluxAnsatzVerify}, we can identify the flux insertion operators or the large gauge transformations for the polynomial shift symmetries Eq.~\eqref{eq.generalPolynomial} as follows: 
\begin{equation}
\label{eq.fluxInsertionMonomialSymmetry}
\begin{split}
\hat{D}_{L,a} = \exp\left( i \frac{2\pi}{L_x} \frac{1}{a!(m+1-a)!}\sum_{\br} x_{\br}^{a} y_{\br}^{m+1-a} \hat{n}_{\br} \right),
\quad\forall\; a=1,\ldots,m+1.
\end{split}
\end{equation}
where the subscript $L$ in $\hat{D}_{L,a}$ refers to ``large" gauge transformation. Note that when $m=1$, this set of operators reduces to the two operators in the first column of Eq.~\eqref{eq.Dij} as expected. The commutation relations between $\hat{T}_x$ and $\hat{D}_{L,a}$ are:
\begin{equation}
\label{eq.generalCommutationOfDandT}
\hat{T}_x \hat{D}_{L,a} \left(T_x\right)^{-1}
=
\hat{D}_{L,a} 
\exp\left( - i \frac{2\pi}{L_x} \frac{1}{a!(m+1-a)!}\sum_{\br} ((x_{\br}+1)^{a}-x_{\br}^{a}) y_{\br}^{m+1-a} \hat{n}_{\br} \right),
\quad\forall\; a=1,\ldots,m+1.
\end{equation}

Hence, if there exists a unique gapped ground state which preserves both the polynomial shift symmetry and translation symmetry, the large gauge transformations must commute with the translation operators. Hence, the phase factor in Eq.~\eqref{eq.generalCommutationOfDandT} should be trivial:
\begin{equation}
\label{eq.generalPolynomialConstraint2}
\begin{split}
&\frac{2\pi}{L_x} \frac{1}{a!(m+1-a)!}\sum_{\br} ((x_{\br}+1)^{a}-x_{\br}^{a}) y_{\br}^{m+1-a} \hat{n}_{\br} \in 2\pi \mathbb{Z},
\quad\forall\; a=1,\ldots,m+1.  \\
\Longleftrightarrow&
\frac{2\pi}{L_x} \frac{1}{a!(m+1-a)!}\sum_{\br} \sum_{s=0}^{a-1} \combinationNumber{a}{s} x_{\br}^s y_{\br}^{m+1-a} \hat{n}_{\br} \in 2\pi \mathbb{Z},
\quad\forall\; a=1,\ldots,m+1.  \\
\end{split}
\end{equation}
which constrains the densities of composite mulitpolar objects of order $\leq m$.

In summary, Eqs.~\eqref{eq.generalPolynomialConstraint1} and \eqref{eq.generalPolynomialConstraint2} impose constraints on the multipoles per unit length in order for an interacting bosonic system to harbor a symmetry-preserving, gapped, and non-degenerate ground state.


\subsection{GSD from fractionalization and planar subsystem symmetries in 3d}
\label{app:C}

In this Appendix, we provide details regarding the ground state degeneracy (GSD) of a three-dimensional gapped system with translation and U(1) planar symmetries, under the added assumption of charge fractionalization. We impose PBC in all three directions and label the planes of the resulting $L_x \times L_y \times L_z$ three-torus as $C_{xy}(z_0)$, $C_{yz}(x_0)$ and $C_{zx}(y_0)$. For instance, $C_{xy}(z_0)$ refers to the $xy$ plane whose $z$-coordinate is located at $z_0$. $C_{yz}(x_0), C_{zx}(y_0)$ are similarly defined. The intersection between $C_{xy}(z_0)$ and $C_{yz}(x_0)$ is denoted $C_{y}(x_0,z_0)$, which is a line in the $y$-direction with $x$ and $z$ coordinates located at $(x_0,z_0)$. Intersections between other planes are analogously defined.

Under our assumption of fractionalization, the planar U(1) subsystem symmetry on each plane supports point-like fractional excitations. Due to the additional assumption of translation invariance, we can assume that the fractional charge on each plane is uniform and equals $p/q$ (in units of $e$), with $p$ and $q$ coprime. Moreover, intersection lines such as $C_x(y_0,z_0)$ can support a composite charge which carries a $p/q$ charge from the plane $C_{xy}(z_0)$ and a $p/q$ charge from the plane $C_{zx}(y_0)$.

For $d=2$ gapped systems with fractionalized excitations, Ref.~[\onlinecite{oshikawasenthil}] showed that the following procedure should commute with the interacting Hamiltonian: (1) a fractionalized excitation and its anti-particle are created from the ground state, (2) the fractionalized excitation is adiabatically transported around a non-trivial cycle of the two-torus, and (3) the pair of excitations annihilate each other, returning the system to the ground state manifold. We now adapt this argument to the case under consideration \textit{i.e.,} a 3d gapped system with planar U(1) subsystem symmetries and fractionalized charges.

Let us denote by $\hat{T}(C_x(y_0,z_0))$ the operator which adiabatically moves a composite charge along the intersection line $C_x(y_0,z_0)$, thereby dragging the charge around a non-trivial cycle in the $x$-direction, anchored at $y=y_0, z=z_0$. We can similarly define operators which move composite charges---pairs of charges coming from two intersecting planes---along any intersection line. The number of such operators is $L_xL_y+L_yL_z+L_zL_x$. Besides moving the fractionalized charges, we can also consider moving the $2\pi$ flux on each plane along the non-trivial cycles on that plane. The operators which implement the adiabatic transport of a unit flux quantum along a non-trivial cycle are denoted:
$$\hat{W}_x(C_{xy}(z)), \hat{W}_y(C_{xy}(z)), \hat{W}_y(C_{yz}(x)), \hat{W}_z(C_{yz}(x)), \hat{W}_z(C_{zx}(y)), \hat{W}_x(C_{zx}(y)).$$
For instance, $\hat{W}_x(C_{xy}(z_0))$ here refers to the operator which moves a $2\pi$ flux along a non-trivial cycle in the $x$-direction on the $xy$ plane with $z=z_0$. 

At the end of each of the adiabatic processes described above, the system remains in the ground state manifold and only accumulates a trivial Aharanov-Bohm phase, such that Hamiltonian commutes with each of the operators $\hat{T}, \hat{W}$. However, these sets of operators satisfy the following non-trivial commutation relations with each other:
\begin{equation}
\begin{split}
\hat{W}_x(C_{xy}(z_0)) \hat{T}(C_{y}(x_0,z_0)) &= e^{-\frac{2\pi i p}{q}} \hat{T}(C_{y}(x_0,z_0)) \hat{W}_x(C_{xy}(z_0)),    \\
\hat{W}_y(C_{xy}(z_0)) \hat{T}(C_{x}(y_0,z_0)) &= e^{-\frac{2\pi i p}{q}} \hat{T}(C_{x}(y_0,z_0)) \hat{W}_y(C_{xy}(z_0)),    \\
\hat{W}_y(C_{yz}(x_0)) \hat{T}(C_{z}(x_0,y_0)) &= e^{-\frac{2\pi i p}{q}} \hat{T}(C_{z}(x_0,y_0)) \hat{W}_y(C_{yz}(x_0)),    \\
\hat{W}_z(C_{yz}(x_0)) \hat{T}(C_{y}(z_0,x_0)) &= e^{-\frac{2\pi i p}{q}} \hat{T}(C_{y}(z_0,x_0)) \hat{W}_z(C_{yz}(x_0)),    \\
\hat{W}_z(C_{zx}(x_0)) \hat{T}(C_{x}(y_0,z_0)) &= e^{-\frac{2\pi i p}{q}} \hat{T}(C_{x}(y_0,z_0)) \hat{W}_z(C_{zx}(x_0)),   \\
\hat{W}_x(C_{zx}(x_0)) \hat{T}(C_{z}(x_0,y_0)) &= e^{-\frac{2\pi i p}{q}} \hat{T}(C_{z}(x_0,y_0)) \hat{W}_x(C_{zx}(x_0)),   \\    
&\quad \forall\;
x_0=1,2,\ldots,L_x,
y_0=1,2,\ldots,L_y,
z_0=1,2,\ldots,L_z.\\
\end{split}
\end{equation}
The GSD generated by this $\mathbb{Z}_q$ algebra is 
\begin{equation}
\mathrm{GSD} = q^{2L_x+2L_y+2L_z-3}
\end{equation}
which grows exponentially with the linear extent of the system.

\end{document}